# Superconductivity in New Iron Pnictide Oxide $(Fe_2As_2)(Sr_4(Mg,Ti)_2O_6)$

Shinya Sato[1,4], Hiraku Ogino[1,4*], Naoto Kawaguchi[1,4], Yukari Katsura[2], Kohji Kishio[1,4], Jun-ichi Shimoyama[1,4], Hisashi Kotegawa[3,4] and Hideki Tou[3,4]

[1]Department of Applied Chemistry, The University of Tokyo, 7-3-1 Hongo, Bunkyo-ku, Tokyo 113-8656, Japan
[2]Magnetic Materials Laboratory, RIKEN, 2-1 Hirosawa, Wako-shi, Saitama 351-0106 Japan
[3]Department of Physics, Kobe University, Kobe 657-8501
[4]JST-TRIP, Sanban-cho, Chiyoda-ku, Tokyo 102-0075, Japan
e-mail address : tuogino@mail.ecc.u-tokyo.ac.jp

## Abstract

We have discovered a new iron pnictide oxide superconductor $(Fe_2As_2)(Sr_4(Mg,Ti)_2O_6)$. This material is isostructual with $(Fe_2As_2)(Sr_4M_2O_6)$ ($M$ = Sc, Cr, V), which were found in previous studies. The structure of this compound is tetragonal with a space group of *P4/nmm* and consists of the anti-fluorite type FeAs layer and perovskite-type block layer. The lattice constants are $a$ = 3.935 Å and $c$ = 15.952 Å for $(Fe_2As_2)(Sr_4MgTiO_6)$. Bulk superconductivity with $T_{c(onset)}$ of ~26 K was observed for a partially Co substituted sample. Moreover, Co-free and Ti-rich samples exhibited higher $T_{c(onset)}$'s above 35 K, which were further enhanced by applying high pressures up to ~43 K.

## Introduction

The discovery of high-$T_c$ superconductivity in LaFeAs(O,F)[1] has triggered researches for development of new iron pnictide superconductors. Until now, several groups of superconductors containing anti-fluorite pnictide or chalcogenide layer have been discovered such as *RE*FeAsO(*RE* = rare earth elements) (abbreviated as 1111)[2],

*AE*$Fe_2As_2$(*AE* = alkali earth metals)[3], LiFeAs[4], and FeSe[5]. In addition, many layered compounds composed of both antifluorite pnictide layer and perovskite-type oxide layer, such as ($Fe_2As_2$)($Sr_3Sc_2O_5$)[6] and ($Fe_2P_2$)($Sr_4Sc_2O_6$)[7], have been successively found. ($Fe_2P_2$)($Sr_4Sc_2O_6$) having $K_2NiF_4$-type perovskite-related oxide layer showed superconductivity at 17 K, which is the highest value among the iron phosphide compounds. Its arsenic relatives ($Fe_2As_2$)($Sr_4M_2O_6$) were also discovered in *M* = Sc, Cr[8,9] and V[10]. Large amount of Ti substitution for ($Fe_2As_2$)($Sr_4Sc_2O_6$) and ($Fe_2As_2$)($Sr_4Cr_2O_6$) were also suggested to be effective for inducing superconductivity[11,12]. Moreover, ($Fe_2As_2$)($Sr_4V_2O_6$) was found to show superconductivity with $T_{c(onset)}$ of ~40 K without intensive carrier doping[10,13] and $T_{c(onset)}$ increased up to 46 K under a pressure of 4 GPa[14]. These facts indicated that this system is a new vein of iron pnictide superconductors.

There are several perovskite oxides called "double perovskite", which have mixed and ordered B-site cations, such as $LaMg_{0.5}Ti_{0.5}O_3$[15] and $LaZn_{0.5}Ti_{0.5}O_3$[16]. Since the $Sr_4M_2O_6$ layer in above layered pnictide oxides have perovskite-type structure, where the *M*-site corresponds to the B-site of perovskite, mixing *M* elements was attempted to develop new layered oxide pnictide compounds in the present study. As a result, new iron arsenide oxide ($Fe_2As_2$)($Sr_4MgTiO_6$) was successfully synthesized. Partial substitution of Co for the Fe-site was found to improve electronic state of this compound, resulting in bulk superconductivity with $T_{c(onset)}$ up to 26 K. This phase was also formed from Co-free and Ti-rich composition as a main phase. The superconducting properties of Co-free and Ti-rich compounds varied with the starting composition and the superconducting transition was observed up to 39 K.

## Experimental

All samples were synthesized by the solid-state reaction method starting from FeAs(3N), CoAs(3N), SrO(2N), MgO(3N), Ti(3N) and $TiO_2$(3N). Nominal compositions were fixed according to the general formula: $((Fe_{1-x}Co_x)_2As_2)(Sr_4(Mg_{1-y}Ti_y)_2O_6)$. Since the starting reagent, SrO, is sensitive to moisture in air, manipulations were carried out under argon atmosphere. Powder mixture of FeAs, SrO, MgO, Ti and $TiO_2$ was pelletized and sealed in evacuated quartz ampoules. Heat-treatments were performed in the temperature range from 1000 to 1250°C for 40 to 72 hours. Phase identification was carried out by X-ray diffraction (XRD) using RIGAKU Ultima-IV diffractometer and intensity data were collected in the $2\theta$ range of 5° - 80° at a step of 0.02° using Cu-*K*α radiation. Silicon powder was used as the internal standard. High-resolution images were taken by a field-emission-type transmission electron microscopy (TEM, JEOL JEM-2010F). Magnetic susceptibility measurement was performed by a SQUID magnetometer (Quantum Design MPMS-XL5s). Electric resistivity was measured by the AC four-point-probe method under fields up to 9 T using Quantum Design PPMS. Electrical resistivity under high pressures was also measured using an indenter cell and Daphne 7474 as a pressure-transmitting medium[17,18]. Applied pressure was estimated from the $T_c$ of the lead manometer.

**Result and discussion**

Figure 1 shows a powder XRD pattern of $(Fe_2As_2)(Sr_4MgTiO_6)$ reacted at 1200°C for 40 h together with a simulation pattern. $(Fe_2As_2)(Sr_4MgTiO_6)$ was obtained as the main phase with a small amount of impurities, such as $SrFe_2As_2$ and FeAs. Although several perovskite oxides with mixed B-site cation is known to have ordered structure accompanying orthorhombic distortion, the XRD pattern of $(Fe_2As_2)(Sr_4MgTiO_6)$ could be indexed as space group of *P4/nmm* and lattice constants were determined to be $a$ = 3.935 Å and $c$ = 15.952 Å. The $a$-axis length was close to that of $(Fe_2As_2)(Sr_4V_2O_6)$ with $a$ = 3.930 Å and $c$ = 15.666 Å[10], while the $c$-axis length is slightly longer.

Figure 2 shows a bright-field TEM image and an electron diffraction pattern taken from [1 -1 0] direction of a $(Fe_2As_2)(Sr_4MgTiO_6)$ crystal. Both TEM image and electron diffraction patterns indicated tetragonal cell with $c/a$ ~ 4.0, which coincided well with a corresponding value analyzed from XRD data. Any satellite spots suggesting superstructure due to ordering of B-site cations were not observed in the electron diffraction patterns.

Powder XRD patterns and lattice constants of $((Fe_{1-x}Co_x)_2As_2)(Sr_4MgTiO_6)$ for $0 \leq x \leq 0.15$ are shown in Fig. 3. Peak intensities of impurity phases did not increase by partial Co-substitution. The $c$-axis length systematically decreased with an increase in Co-substitution level, $x$, while the $a$-axis lengths were almost unchanged.

Temperature dependences of ZFC and FC magnetization for $((Fe_{1-x}Co_x)_2As_2)(Sr_4MgTiO_6)$ with $0 \leq x \leq 0.15$ measured under 1 Oe are shown in Fig. 4. The Co-free sample showed diamagnetism due to superconductivity with $T_{c(onset)}$ below 10 K and its volume fraction was only 1.5 %. On the other hand, large diamagnetism were observed in the Co-doped samples, and $T_{c(onset)}$ were 22 K, 24 K, 16 K for $x$ = 0.05, 0.1 and 0.15, respectively. Figure 5 shows temperature dependence of resistivity for $((Fe_{1-x}Co_x)_2As_2)(Sr_4MgTiO_6)$ with $0 \leq x \leq 0.15$. The metallic behaviors were observed in the normal state resistivity for all samples and $T_{c(onset)}$ were ~ 10 K, ~ 26 K, ~ 26 K and 18 K for $x$ = 0, 0.05, 0.1 and 0.15, respectively. Zero resistivity was confirmed at 2 K, 15 K, 11 K and 9 K for $x$ = 0, 0.05, 0.1 and 0.15, respectively. It should be noted that this is the first example of positive Co-substitution effect on superconducting properties of the $(Fe_2As_2)(Sr_4M_2O_6)$ system.

Figure 6 shows powder XRD patterns of $(Fe_2As_2)(Sr_4(Mg_{1-y}Ti_y)_2O_6)$ reacted at 1200°C for 40 h. $(Fe_2As_2)(Sr_4(Mg_{1-y}Ti_y)_2O_6)$ phase was obtained for $0.5 \leq y \leq 0.7$ as main phase, while peaks due to $SrFe_2As_2$, $SrTiO_3$ and other minor impurities increased with increasing $y$. On the other hand, Mg-rich compound was not formed judging from the abrupt increase in peak intensities of impurities and small changes in lattice constants. Any pnictide oxide phase could not be found in $y$ = 1.0. For samples with $0.5 \leq y \leq 0.7$, the $c$-axis length systematically decreased with an increase of $y$, while it was difficult to calculate lattice constant for $y$ = 0.8 due to overlapping of diffraction peaks with impurities. This indicates that the compositional range of solid-solution exists towards Ti-rich composition.

Figure 7 shows backscattered electron images of a sample with a nominal composition of $(Fe_2As_2)(Sr_4(Mg_{0.3}Ti_{0.7})_2O_6)$. Three different grains were observed and EDX spectrum for each grains suggested that white, gray and black grains corresponded to $SrFe_2As_2$, $(Fe_2As_2)(Sr_4(Mg,Ti)_2O_6)$, and $SrTiO_3$ or $Sr_2TiO_4$, respectively.

Temperature dependences of ZFC and FC magnetization for $(Fe_2As_2)(Sr_4(Mg_{1-y}Ti_y)_2O_6)$ of $0.5 \leq y \leq 1$ measured under 1 Oe are shown in Fig. 8. $T_{c(onset)}$ increased with an increase of $y$ up to 0.8, while superconducting volume fractions indicated by ZFC magnetization at low temperatures were large for samples with $x$ = 0.6 and 0.7. In particular, samples with $y$ = 0.55, 0.6 and 0.7 exhibited large diamagnetism suggesting bulk superconductivity. The highest $T_{c(onset)}$ 37 K was achieved by $(Fe_2As_2)(Sr_4(Mg_{1-y}Ti_y)_2O_6)$ with $y$ = 0.8. It should be noted that we confirmed that $SrFe_2As_2$ did not show superconductivity even by doping of Ti, Mg or co-dopings of Ti and O or Mg and O.

Figure 9 shows temperature dependence of resistivity for $(Fe_2As_2)(Sr_4(Mg_{1-y}Ti_y)_2O_6)$ of $0.5 \leq y \leq 0.8$. The metallic behaviors were observed in the normal state resistivity for all samples, and $T_{c(onset)}$ was ~ 33 K, ~ 34 K, ~36 K and 39 K for $y$ = 0.55, 0.6, 0.7 and 0.8, respectively. Zero resistivity was achieved at 14 K, 18 K, 22 K and 17 K for $y$ = 0.55, 0.6, 0.7 and 0.8, respectively. These broad superconducting transitions indicate poor grain connectivity of the samples or inhomogeneous cation compositions.

The $T_c$ increased systematically with an increase of $y$ up to 39 K. In addition, superconducting volume fraction also increased with increasing $y$. Superconductivity in Co-doped and Ti-rich samples indicates that electron doping is also effective in this compound as in the 1111 system. Very small superconducting volume fraction of the $y$ = 0.5 sample indicates that this compounds doesn't exhibit superconductivity intrinsically without intensive doping.

Figure 10 shows temperature dependence of resistances measured under various pressures for $(Fe_2As_2)(Sr_4(Mg_{0.3}Ti_{0.7})_2O_6$. The $T_{c(onset)}$ increases with an increase of the pressure similarly in the case of $(Fe_2As_2)(Sr_4V_2O_6)$[14] and reaches ~43 K at 4.2 GPa. The $T_{c(onset)}$ first increases largely by applied pressure with $dT_c/dP$ = 2.5 K/GPa and it seems to saturate at approximately 4 GPa.

## Conclusions

A new layered iron pnictide oxide $((Fe_{1-x}Co_x)_2As_2)(Sr_4(Mg_{1-y}Ti_y)_2O_6)$ was synthesized and its physical properties were characterized. This material has alternate stacking of anti-fluorite $Fe_2As_2$ and perovskite-type $Sr_4(Mg,Ti)_2O_6$ layers. This compound exhibits superconductivity by partial substitution of Co for the Fe-site, and $T_{c(onset)}$ is 26 K at $x$ = 0.05. This phase is also obtained as a main phase in the Ti-rich starting composition. In resistivity measurements, superconducting transition is observed up to 39 K under the ambient pressure, and it was increased up to ~43 K by applying a high pressure of 4.2 GPa. These facts indicate that superconductivity of this compound is induced by the carrier control of the $Fe_2As_2$ layer.

Ackowledgement

This work was partly supported by Grant-in-Aid for Young Scientists (B) no. 21750187, 2009, supported by the Ministry of Education, Culture, Science and Technology (MEXT) as well as inter-university Cooperative Research Program of the Institute for Materials Research, Tohoku University.

## Figure captions

Figure 1. Powder XRD and simulation patterns of $(Fe_2As_2)(Sr_4MgTiO_6)$.

Figure 2. Bright-field TEM image and corresponding electron diffraction pattern of a $(Fe_2As_2)(Sr_4MgTiO_6)$ crystal viewed from [1 -1 0] direction.

Figure 3. Powder XRD patterns and lattice constants of $((Fe_{1-x}Co_x)As_2)(Sr_4MgTiO_6)$ for $0 \leq x \leq 0.15$.

Figure 4. Temperature dependence of ZFC and FC magnetization curves of $((Fe_{1-x}Co_x)As_2)(Sr_4MgTiO_6)$ bulk samples measured under 1 Oe. Close-up of magnetization curves for $((Fe_{1-x}Co_x)As_2)(Sr_4MgTiO_6)$ is shown in the inset.

Figure 5. Temperature dependences of resistivity for the $((Fe_{1-x}Co_x)As_2)(Sr_4MgTiO_6)$ bulks at $0 < T < 50$ K. Temperature dependences of resistivity at $0 < T < 300$ K is shown in the inset.

Figure 6. Powder XRD patterns and lattice constants of $(Fe_2As_2)(Sr_4(Mg_{1-y}Ti_y)_2O_6)$ for $0.4 \leq y \leq 1$.

Figure 7. Backscattered electron image of $(Fe_2As_2)(Sr_4(Mg_{0.3}Ti_{0.7})_2O_6)$.

Figure 8. Temperature dependence of ZFC and FC magnetization curves of $(Fe_2As_2)(Sr_4(Mg_{1-y}Ti_y)_2O_6)$ bulk samples measured under 1 Oe. Close-up of the curves are shown in the inset.

Figure 9. Temperature dependences of resistivity for the $(Fe_2As_2)(Sr_4(Mg_{1-y}Ti_y)_2O_6)$ bulks at $0 < T < 50$ K. Temperature dependences of resistivity at $0 < T < 300$ K is shown in the inset.

Figure 10. Temperature dependences of resistance under various pressure for the $(Fe_2As_2)(Sr_4(Mg_{0.3}Ti_{0.7})_2O_6)$ bulks at $20 < T < 60$ K. Temperature dependences of resistivity at $0 < T < 300$ K is shown in the inset.

## References

[1] Kamihara Y, Watanabe T, Hirano M and Hosono H, 2008 *J. Am. Chem. Soc.* **130** 3296.

[2] Quebe P, Terbüchte L, W. Jeitschko W, 2000 *J. Alloys Compd.* **302** 70.

[3] Rotter M, Tegel M and Johrendt D, 2008 *Phys. Rev. Lett.* **101** 107006.

[4] Pitcher M, Parker D, Adamson P, Herkelrath S, Boothroyd A, Ibberson R, Brunelli M and Clarke S, 2008 *Chem. Commun.* **45** 5918

[5] Hsu F.C, Luo J.Y, The K.W, Chen T.K, Huang T.W, Wu P.M, Lee Y.C, Huang Y.L, Chu Y.Y, Yan D.C and Wu M.K, 2008 Proc. *Natl. Acad. Sci. U.S.A.* **105** 14262..

[6] Zhu X, Han F, Mu G, Zeng B, Cheng P, Shen B, Wen H-H, 2009 *Phys. Rev. B* **79** 024516.

[7] Ogino H, Matsumura Y, Katsura Y, Ushiyama K, Horii S, Kishio K and Shimoyama J. 2009 *Supercond. Sci. Technol.* **22** 075008.

[8] Ogino H, Katsura Y, Horii S, Kishio K and Shimoyama J, 2009 *Supercond. Sci. Technol*. **22** 085001.

[9] Tegel M, Hummel F, Lackner S, Schellenberg I, Pöttgen R, and Johrendt D, 2009 arXiv : Condmat / 0904.0479(unpublished).

[10] Zhu X, Han F, Mu G, Cheng P, Shen B, Zeng B and Wen H-H, 2009 *Phys. Rev. B* **79** 220512(R).

[11] Chen G.F, Xia T.L, Yang H.X, Li J.Q, Zheng P, Luo J.L and Wang N.L, 2009 *Supercond. Sci. Technol.* **22** 072001.

[12] Zhu X, Han F, Mu G, Cheng P, Shen B, Zeng B and Wen H-H, 2009 arXiv : Condmat / 0904. 0972.

[13] Han F, Zhu X, Mu G, Cheng P, Shen B, Zeng B and Wen H-H, 2009 arXiv : Condmat / 0910. 1537.

[14] Kotegawa H, Kawazoe T, Tou H, Murata K, Ogino H, Kishio K, Shimoyama, 2009 arXiv : Condmat / 0908. 1469(unpublished).

[15] Lee D.Y, Yoon S-J, Yeo J.H, Nahm S, Paik J.H, Whang K-C and Ahn B.G, 2000 *J. Mater. Sci. Lett.* **19** 131-134

[16] Ubic R, Hu Y and Abrahams I, 2006 *Acta Crystallogr. B* **62** 521-529.

[17] Kobayashi T.C, Hidaka H, Kotegawa H, Fujiwara K, and Eremets M.I, 2007 *Rev. Sci. Instrum*. **78** 023909.

[18] Murata K, Yokogawa K, Yoshino H, Klotz S, Munsch P, Irizawa A, Nishiyama M, Iizuka K, Nanba T, Okada T, Shiraga Y, and Aoyama S, 2008 *Rev. Sci. Instrum*. **79** 085101.

Figure 1

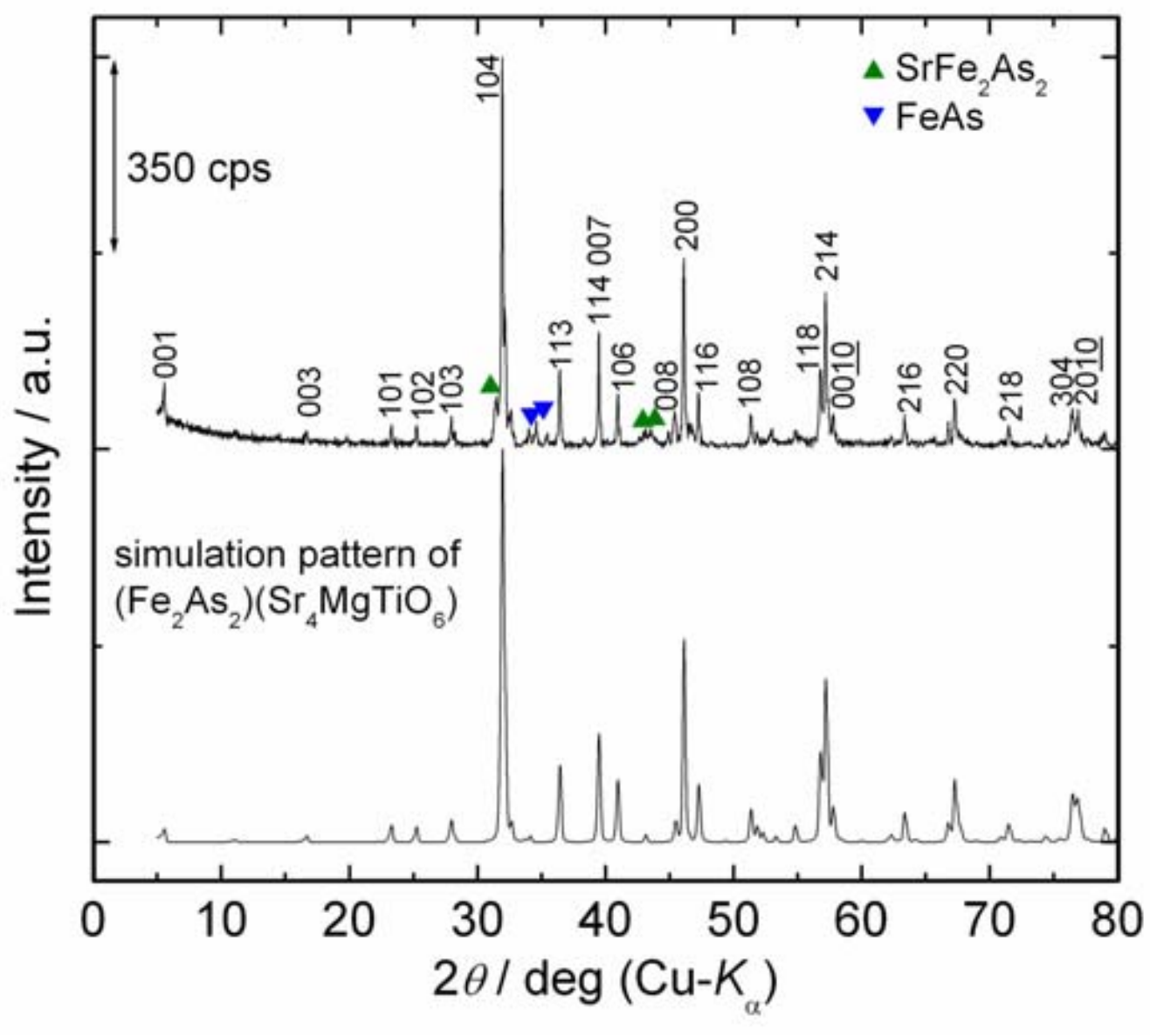

Figure 2

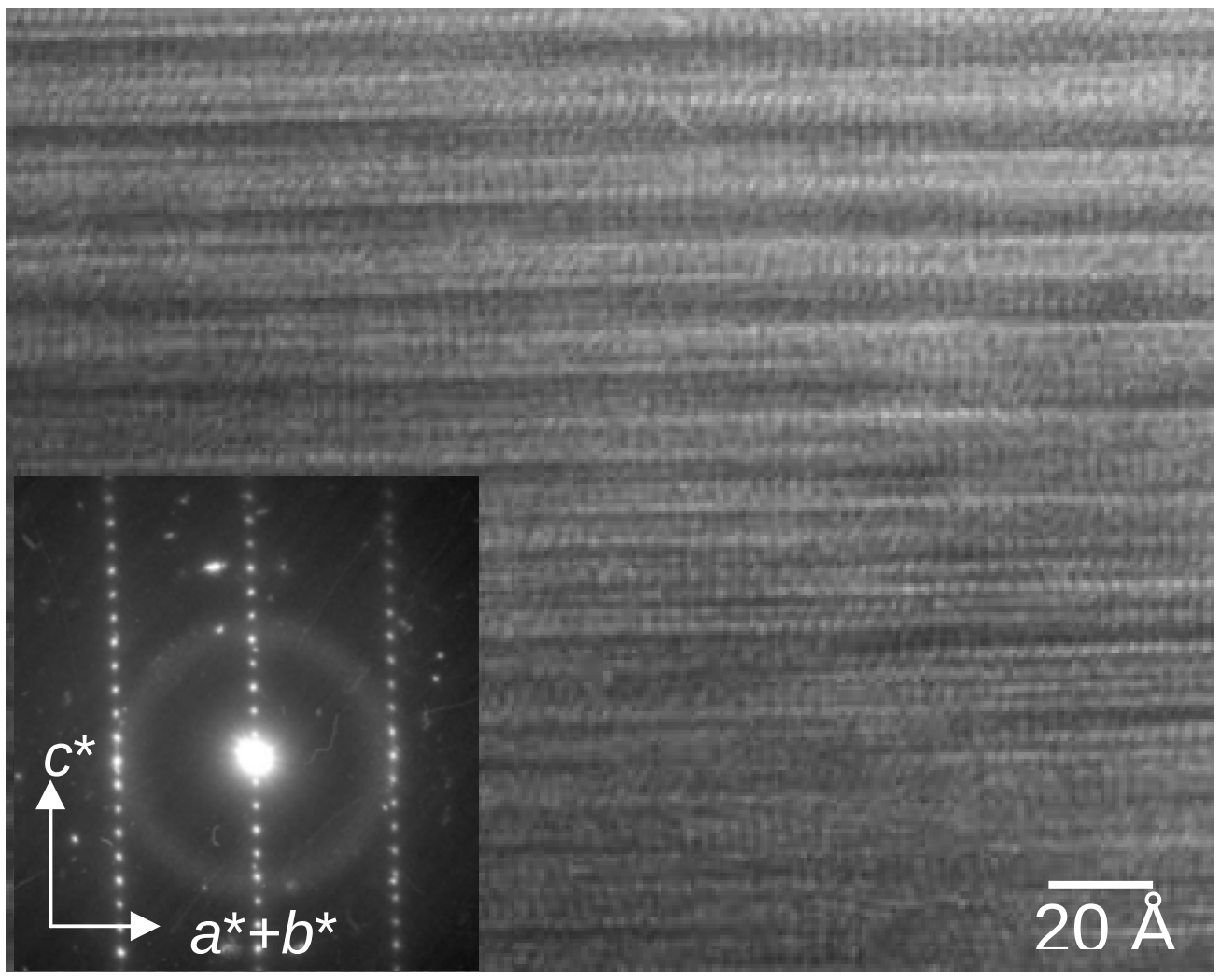

Figure 3

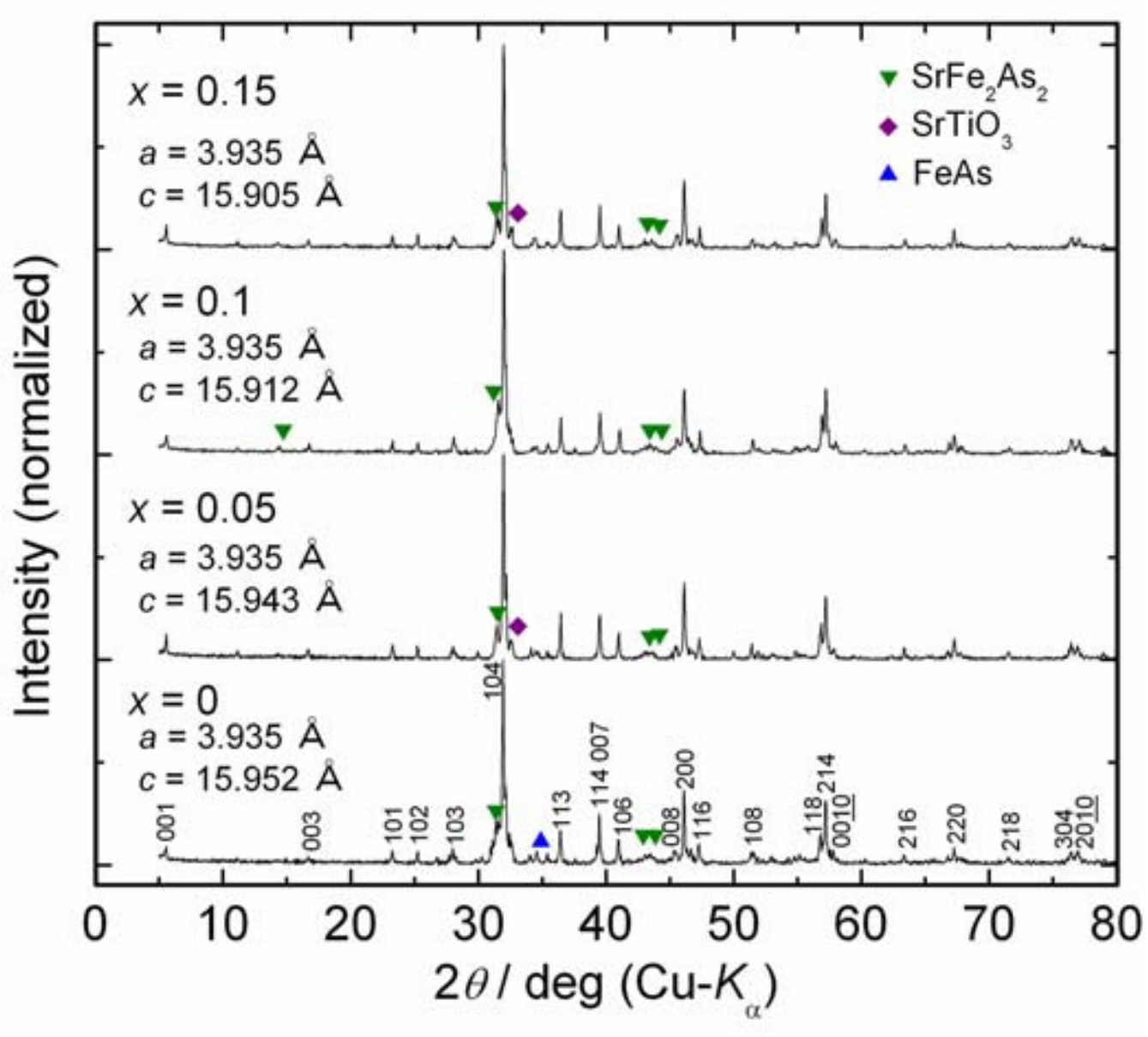

Figure 4

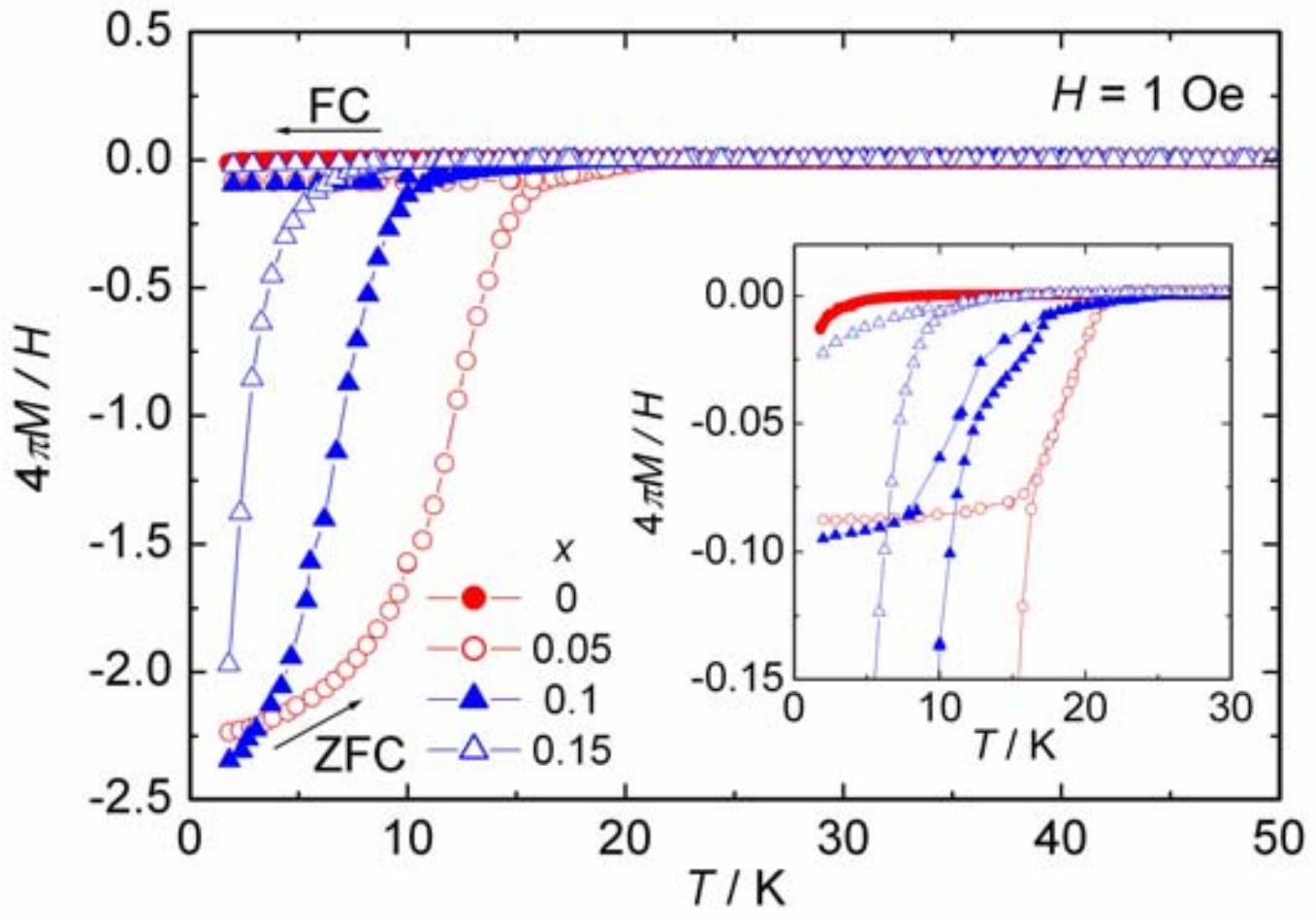

Figure 5

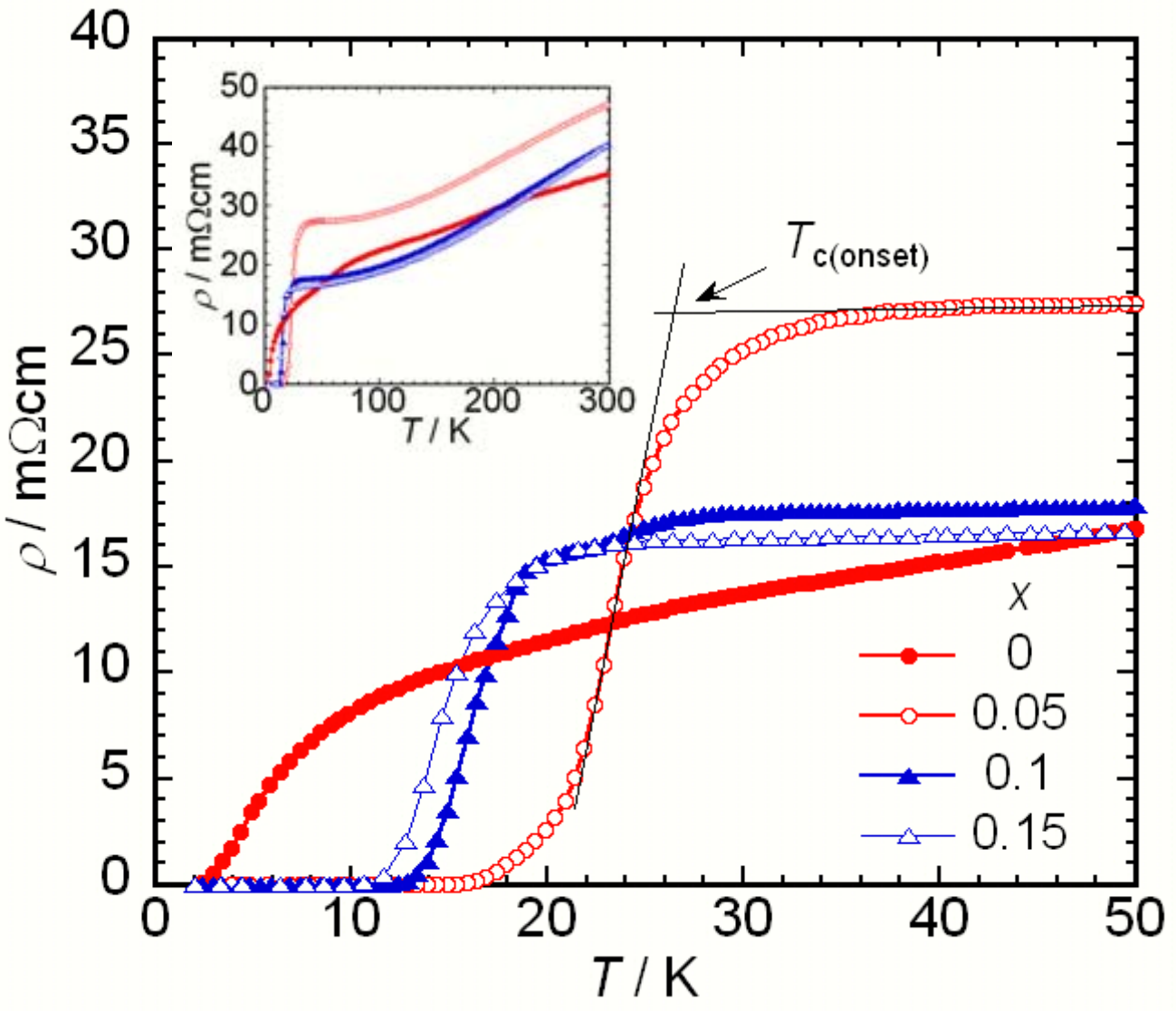

Figure 6

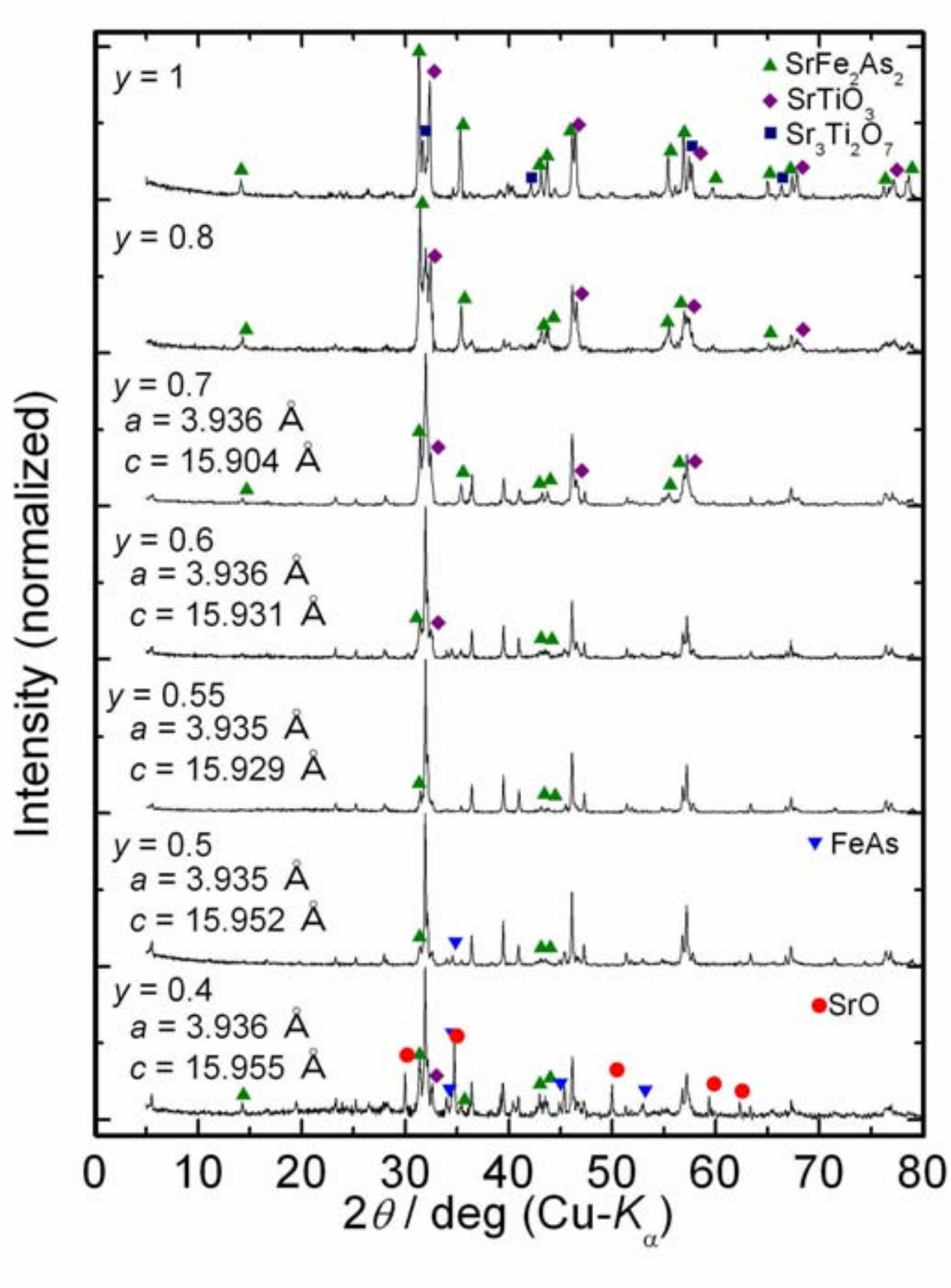

Figure 7

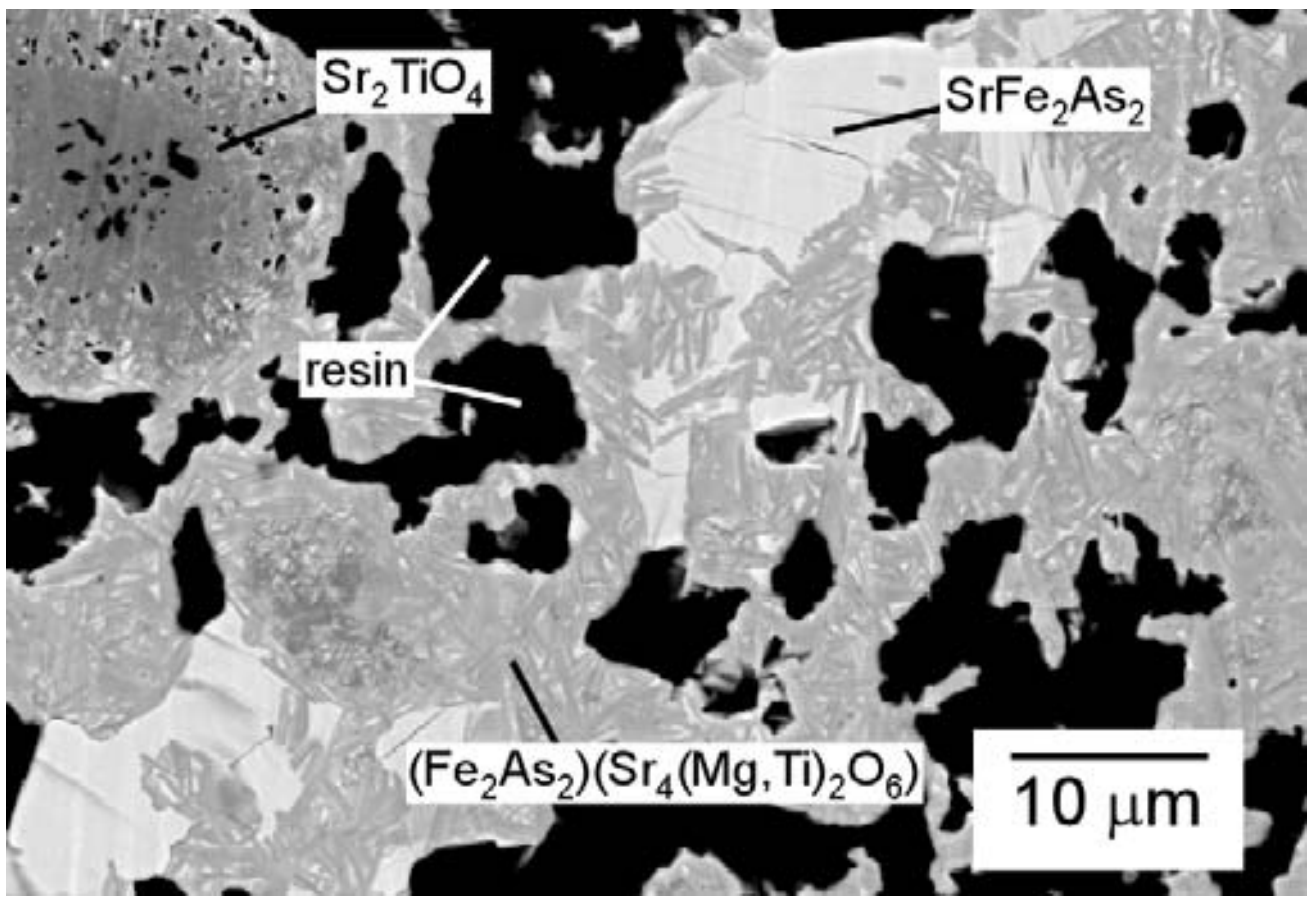

Figure 8

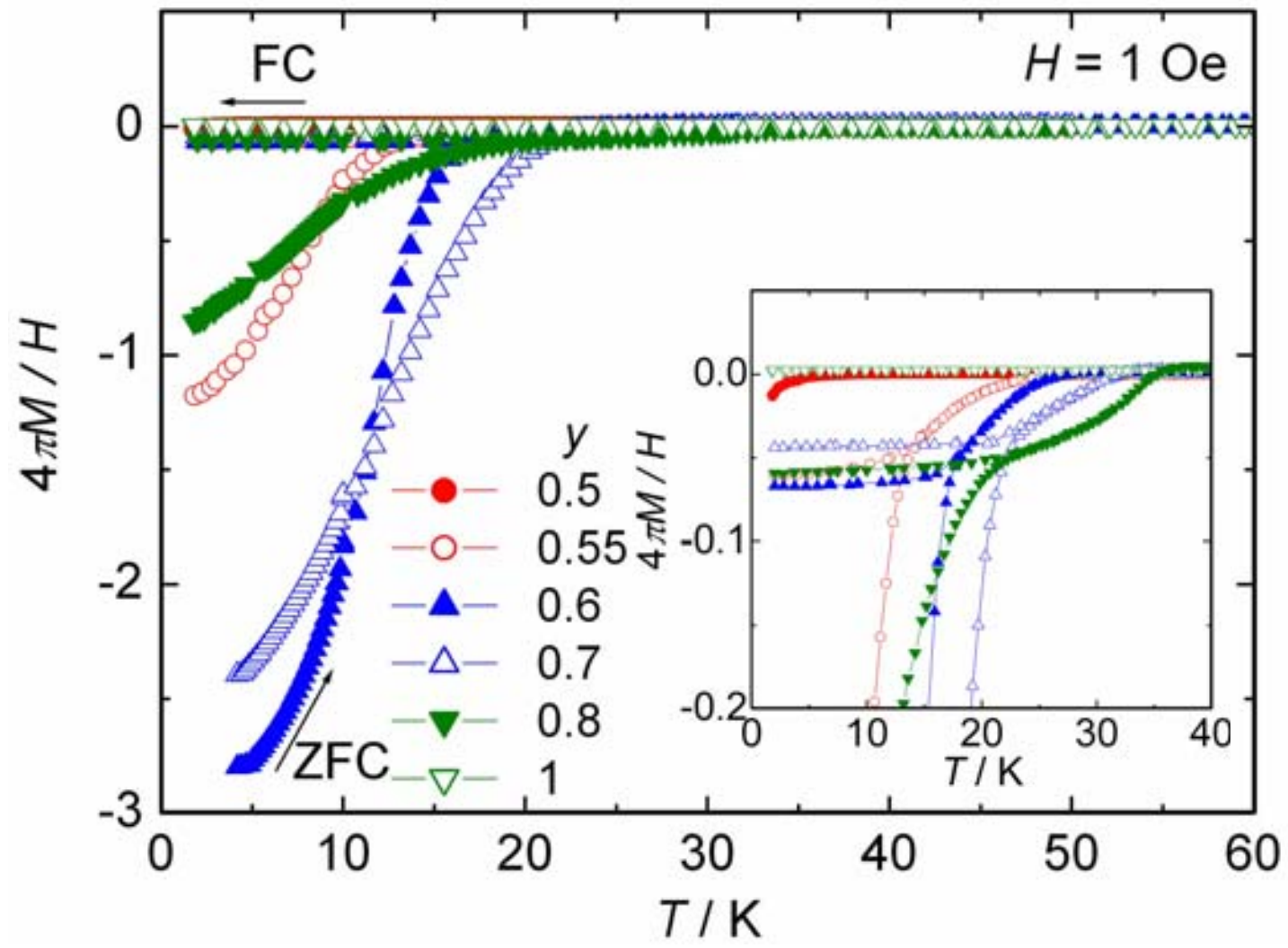

Figure 9

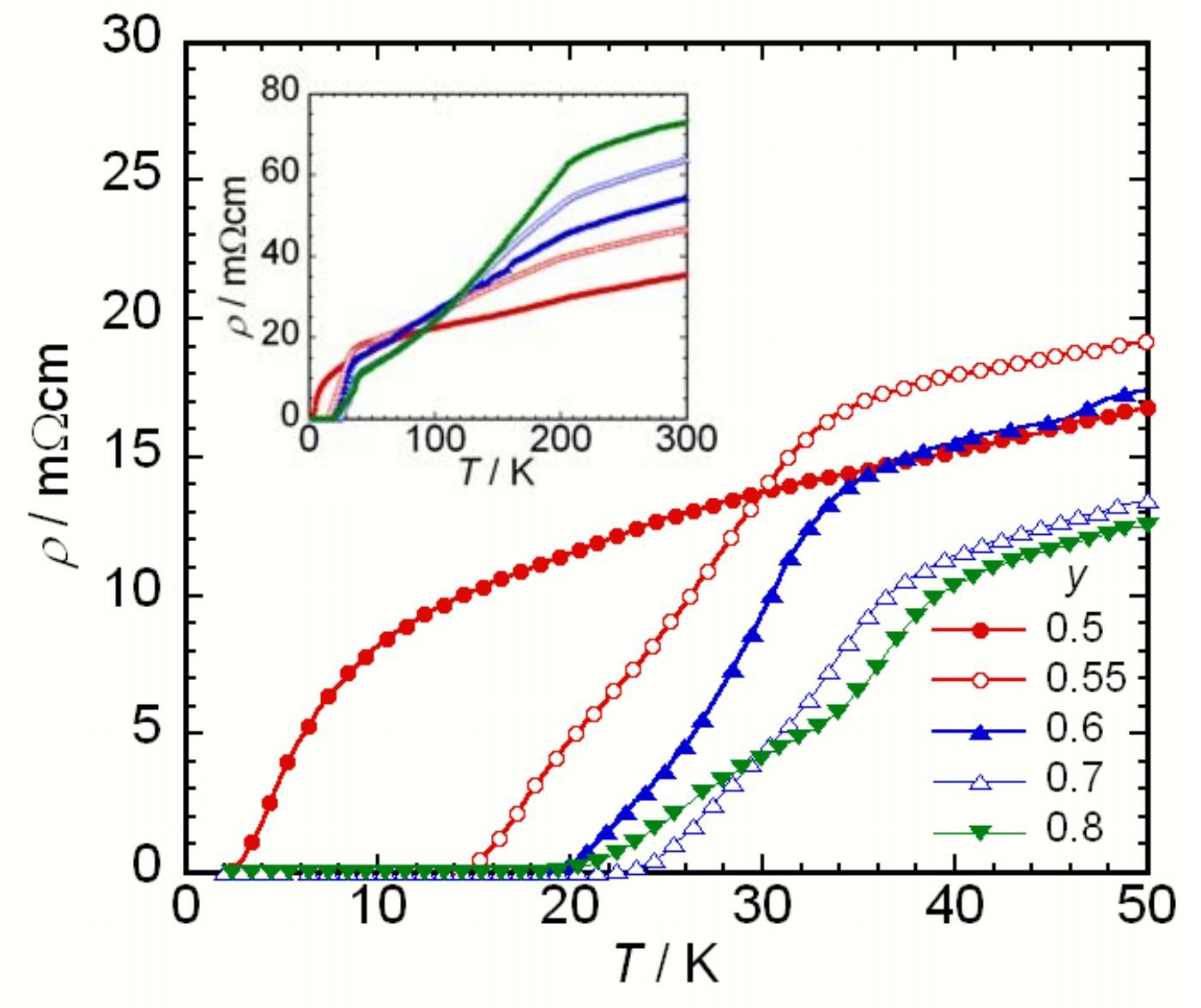

Figure 10

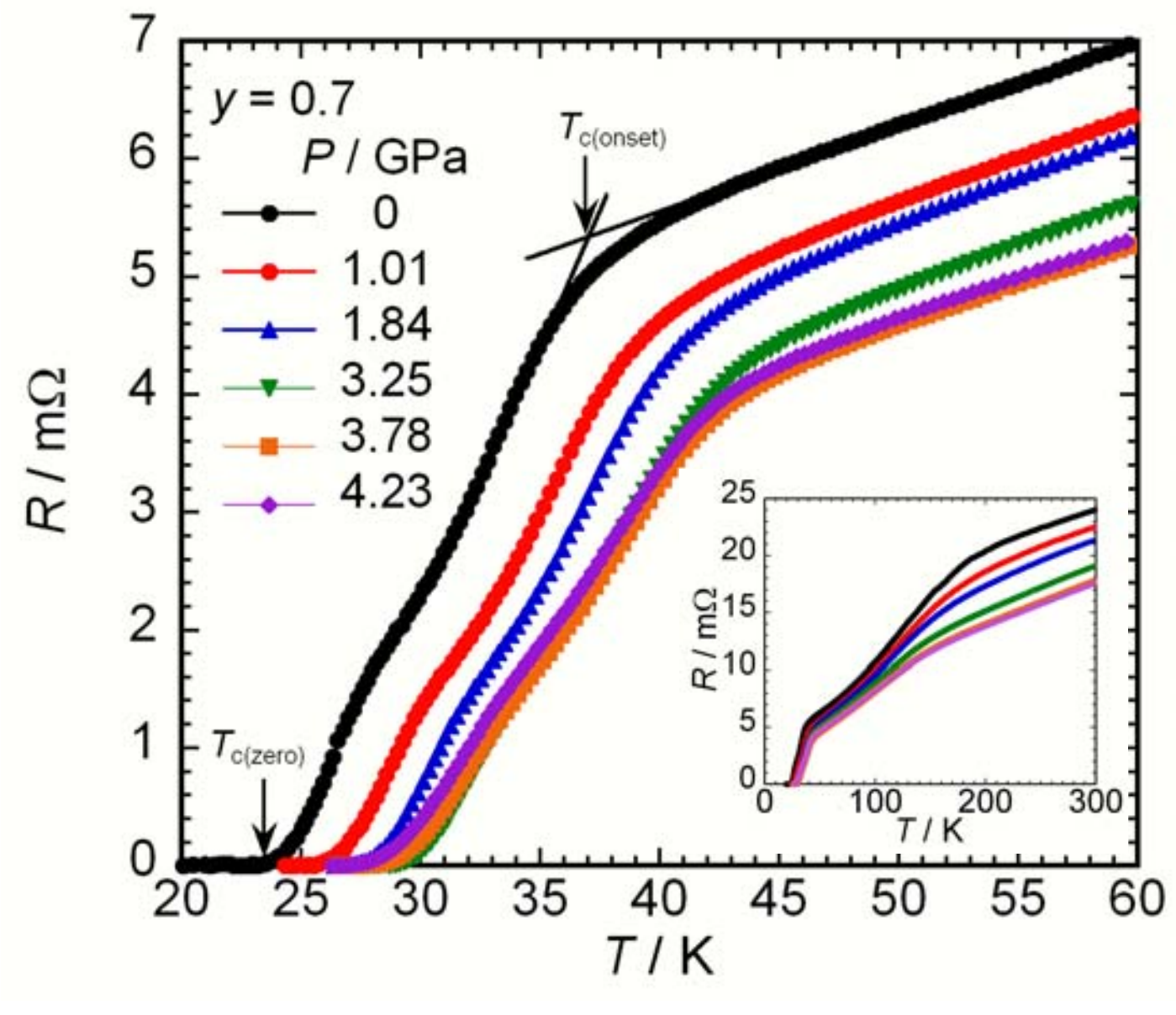